\documentclass[twocolumn,prl,aps,superscriptaddress,showpacs,amsmath,amssymb,floatfix]{revtex4-1}

\usepackage{bm}
\usepackage{epsfig}
\usepackage{graphicx}

\begin{document}

\title{Exact analysis on the singularity of the joint density of states and its relationship to the quasiparticle interference}

\author{Qiang Han}
\email{hanqiang@ruc.edu.cn}
\affiliation{Department of Physics, Renmin University of China,
Beijing, China}

\author{Dan-Bo Zhang}
\affiliation{Department of Physics, Renmin University of China,
Beijing, China}

\author{Z. D. Wang}
\affiliation{Department of Physics and Center of Theoretical and
Computational Physics, The University of Hong Kong, Pokfulam Road,
Hong Kong, China}

\date{\today}
\begin{abstract}

Singularities of the joint density of states (JDOS) and Fourier-transformed local density of states (FT-LDOS) correspond to the hot spots in quasiparticle interference  patterns. In this paper the singularity of JDOS is analyzed exactly, with three types of singularities being classified. In particular, the third type of singularities are found exactly to be envelopes of the contours of constant energy. Remarkably, we show that JDOS and FT-LDOS have the same singular points. Approaching to the singular points, both quantities diverge complementarily in an inverse-square-root manner if the joint curvature is nonzero. The relative magnitude of divergence is governed by the joint curvature as well as the product of the quasiparticle velocities. If the joint curvature of certain singularity is zero, the divergence has a higher order than $-1/2$. 

\end{abstract}

\maketitle

\section{Introduction}

Fifteen yeas ago Fourier-transformed (FT) scanning tunneling microscopy/spectroscopy (STM/STS) first demonstrated its potential as a supplementary technique to probing the surface band structure of metals~\cite{Sprunger}. It is astonishing that the intrinsic momentum-space electronic structure can be reflected by imperfections and obtained by a real-space measurement. Since then the FT-STS technique has been widely applied to study the Fermi surface and energy gap structure in $d$-wave cuprates\cite{Hoffman,McElroy03nature,Fischer} and iron-based superconductors~\cite{Hanaguri2010,Allan} as well as the surface bands of topological insulators \cite{TZhang,Roushan,Beidenkopf} by imaging the modulation of local density of states (LDOS). The modulation is caused by the interference of quasiparticles scattered elastically off imperfection on the material surface. If the quantum interference is between quasiparticles scattered from $\mathbf{k}_i$ to $\mathbf{k}_f$ states on the contours of constant energy (CCE), then the LDOS modulation with wave vectors $\mathbf{q}=\mathbf{k}_f-\mathbf{k}_i$ appears. Based on the quantum transition theory it was argued \cite{Hoffman,McElroy03nature} that the intensity of modulation with wave vector $\mathbf{q}$ is proportional to the joint density of states (JDOS) of the initial and final states, i.e.~$|\nabla E_{\mathbf{k}_i}|^{-1}|\nabla E_{\mathbf{k}_f}|^{-1}$. In $d$-wave cuprates due to the strong anisotropy of the energy dispersion of Bogoliubov quasiparticles, the CCE are banana shaped with large density of states
on their tips. Therefore the scattering of quasiparticles from tip to tip on the CCE are strengthened according to the JDOS picture, which interprets the main findings about the quasiparticle interference (QPI) pattern in high-$T_c$ superconductors.

The relevance of the JDOS picture to FT-STS has been suggested ~\cite{Markiewicz} theoretically and checked experimentally~\cite{McElroy,Hashimoto,Hoffman11}. Numerical~\cite{Allan,DBZhang,QHan} and approximately analytical~\cite{DBZhang,QHan} works show that JDOS surely can repeat the QPI pattern observed by FT-STS and match the observed octet vectors quite well. However, the main inconsistency of the JDOS interpretation is that the QPI pattern actually corresponds to the FT-LDOS. JDOS and FT-LDOS are obviously different physical quantities and the matching between them is rather curious.

Under the Dirac cone approximation, Refs.~[\onlinecite{QHan}] and [\onlinecite{DBZhang}] showed analytically that JDOS and FT-LDOS diverge on the same singular boundary on which the octet vector is located. These findings gave a clue on the validity of JDOS picture in explaining the QPI pattern. That is, the hot spots in the QPI pattern observed by FT-STS possibly correspond to the singularities where both JDOS and FT-LDOS strongly diverge. However these works are limited in that (1) it is only applicable to $d$-wave superconductors in low energy limit where the Dirac cone approximation is valid, (2) of all the octet vectors only the analytic result for $\mathbf{q}_1$ can be given and (3) one need to assume that the dominant impurities are nonmagnetic to highlight $\mathbf{q}_1$ from other points on the boundary. If the scatterer is of magnetic or pair-breaking type, $\mathbf{q}_1$ can even become a dark spot according to the results under the Dirac cone approximation\cite{DBZhang}. Actually there are various types of scatterers on the surface of high-$T_c$ cuprates and even a single impurity can scatter Bogoliubov quasiparticles in a complicated and composite manner. (4) The JDOS arguments seem to only applicable to superconductors with strong anisotropy in the energy dispersion. 

The motivation of this paper is (1) to give a concrete theoretical foundation for the JDOS picture by exploring the singularities of JDOS and FT-LDOS exactly, and (2) to
make the JDOS picture also applicable to more general systems.
The main results of this paper are as follows. (1) The singularities connect two momenta at which the quasiparticle group velocities are (anti)parallel to each other.
(2) For any given quasiparticle energy, the singularities form closed curves which are envelopes of the one-parameter families of CCE curves.
(3) JDOS exhibits power-law divergent behavior with exponent $-1/2$ if the singular point is approached from one side. The coefficient of the divergent term is governed by the inverse of joint curvature as well as the quasiparticle velocities. (4) If the joint curvature is zero the power-law exponent is higher. (5) The FT-LDOS shares the same singularities of JDOS and the divergent behavior is also the same except that it diverge if the singular point is approached from the other side.

The main results of our paper are given by Eqs.~(\ref{JDOS}), (\ref{jacobi}), (\ref{sing}) and (\ref{j_curvature}). Some special cases are discussed to show how to make use of these results.
In the following section we first study the JDOS which is easier than FT-LDOS to reveal their singularities.

\section{Singularities of JDOS}

The JDOS, for studying the QPI pattern on the surface of metals, superconductors, topological insulators and so on, is defined as,
\begin{equation}
    \label{JDOSdef}
    J(\mathbf{p},\omega) = \int F_{\mathbf{k},\mathbf{k+p}} \delta(\omega-E_\mathbf{k+p}) \delta(\omega-E_\mathbf{k}) dk_xdk_y,
\end{equation}
where $E_\mathbf{k}=E(k_x,k_y)$ is the energy dispersion of quasiparticles on the surface. $F_{\mathbf{k},\mathbf{k+p}}$ is a regular function and takes into account the effects of coherent factors~\cite{Hanaguri2009,Coleman,DBZhang} if the quasiparticle is a superposition of an electron and a hole, or phase factors if the quasiparticle is a superposition of a spin-up and spin-down electrons.

We first change the variables from $(k_x,k_y)$ to $(E,E^\prime)$ with
$E=E_\mathbf{k}$, $E^\prime=E_\mathbf{k+p}$ and the double integral in Eq.~(\ref{JDOSdef}) is transformed step by step to,
\begin{equation}
\begin{aligned}
    J(\mathbf{p},\omega) &= \int F_{\mathbf{k},\mathbf{k+p}} \frac{\delta(\omega-E^\prime) \delta(\omega-E)}{\left|\frac{\partial (E_\mathbf{k}, E_\mathbf{k+p})}{\partial (k_x k_y)}\right| } dEdE^\prime \\
        &= \sum_{\mathbf{k}_0}  F_{\mathbf{k}_0,\mathbf{k}_0+\mathbf{p}} \left|\frac{\partial (E_\mathbf{k}, E_\mathbf{k+p})}{\partial (k_x k_y)} \right|^{-1}_{\mathbf{k}=\mathbf{k}_0}, \\
        &= \sum_{\mathbf{k}_0} \frac{F_{\mathbf{k}_0,\mathbf{k}_0+\mathbf{p}}}{\left| \mathbf{v}_{\mathbf{k}_0}\times \mathbf{v}_{\mathbf{k}_0+\mathbf{p}}\right|},
\end{aligned}
\label{JDOS}
\end{equation}
where $\mathbf{v}_\mathbf{k}=\nabla_\mathbf{k} E_\mathbf{k}$ the group velocity of the quasiparticle and the sum is over all $\mathbf{k}_0$'s in the real solution set $\mathbb{S}(\mathbf{p},\omega)$ of the following two equations,
\begin{eqnarray}
    E_{\mathbf{k}_0}=\omega, \label{CCE1} \\
    E_{\mathbf{k}_0+\mathbf{p}}=\omega. \label{CCE2}
\end{eqnarray}
It can be seen that Eq.~(\ref{CCE1}) and (\ref{CCE2}) determine two sets of CCE with the latter is just the former translated by $-\mathbf{p}$. Geometrically $\mathbf{k}_0$'s  are all the intersection points of the two CCE. As shown in Eq.~(\ref{JDOS}), the point $(\mathbf{p},\omega)$ in the 3D momentum-energy space is a singularity of JDOS if there exists at least one $\mathbf{k}_0 \in \mathbb{S}(\mathbf{p},\omega)$ satisfies~\cite{note1},
\begin{equation}
    \label{jacobi}
    \mathbf{v}_{\mathbf{k}_0} \times \mathbf{v}_{\mathbf{k}_0+\mathbf{p}} = 0.
\end{equation}

According to Eq.~(\ref{JDOS}), (\ref{CCE1}), (\ref{CCE2}) and (\ref{jacobi}), we can classify three types of singularities of JDOS. The first one is the von-Hove singularities (vHS) which are also the well-known singularities of the density of states (DOS), i.e.~$D(\omega) = \int \delta(\omega-E_\mathbf{k}) d\mathbf{k} = \int_{E_\mathbf{k}=\omega} {|\mathbf{v}_\mathbf{k}|}^{-1} dl$. The vHS of $D(\omega)$ occur at $\omega_\text{cr}$ for which the CCE crossing the critical points $\mathbf{k}_\text{cr}$ with $\mathbf{v}_{\mathbf{k}_\text{cr}}=0$. Similarly  as indicated by Eq.~(\ref{JDOS}) the vHS of $J(\mathbf{p},\omega)$ also occur at the same $\omega_\text{cr}$ of the DOS if $\mathbf{k}_0$ or $\mathbf{k}_0+\mathbf{p}$ coincide with the critical points. Corresponding to $\omega_\text{cr}$ the singularities of JDOS in the momentum space is given by the implicit function $E_{\mathbf{k}_\text{cr}\pm\mathbf{p}}=\omega_\text{cr}$.

The second type is called perfect-nesting singularities, which means that there are uncountably infinite number of $\mathbf{k}_0$ satisfying Eq.~(\ref{CCE1}) and (\ref{CCE2}) for certain vectors $\mathbf{p}$ and $\omega$. Geometrically this case means that the two CCE have finite segment in common, and thereby Eq.~(\ref{jacobi}) is naturally satisfied on the common arc since $\mathbf{v}_\mathbf{k}$ is normal to the CCE. The perfect-nesting type of singularities is quite special and occurs accidentally in real materials except for the trivial self-nesting case with $\mathbf{p}=0$. Actually this type is the limiting case of the third type, i.e.~the nearly-nesting type, of singularities. In the following, we will elaborate on the third type of singularities, which exists inevitably in real systems and therefore is more general and physically more relevant.

If $(\mathbf{p},\omega)$ is a third type of singularities of JDOS, then the coupled Eqs.~(\ref{CCE1}), (\ref{CCE2}) and (\ref{jacobi}) can be satisfied under the conditions: (i) the two CCE do not cross any critical point, (ii)
the number of $\mathbf{k}_0$, i.e.~the number of crossing points of the two CCE, is finite. Geometrically this means that the two CCE are tangent to, or in other words touching with, each other at certain point $\mathbf{k}_0$ as indicated by Eq.~(\ref{jacobi}). Furthermore the tangent point is actually the degenerate root of Eqs.~(\ref{CCE1}) and (\ref{CCE2}), whose multiplicity is an integer greater than $2$. This multiplicity can be assigned to every singular point $(\mathbf{p},\omega)$ to characterize the degree of nesting between the two CCE at $\mathbf{k}_0$ and $\mathbf{k}_0+\mathbf{p}$, and we denote it by $M(\mathbf{p},\omega)$. The above-mentioned second type, i.e.~the perfect-nesting type, of singularities corresponds to $M(\mathbf{p},\omega)=\infty$.
The JDOS shows power-law divergence when any singularity $(\mathbf{p},\omega)$ is approached and the exponent is $M(\mathbf{p},\omega)^{-1}-1$ as we reveal in the next
section for the $M=2$ cases. 

After canceling $\mathbf{k}_0$ of the coupled equations (\ref{CCE1}), (\ref{CCE2}) and (\ref{jacobi}) formally, one can in principle obtain equation for these non-trivial singularities of $J(\mathbf{p},\omega)$. For any given $\omega$ these singularities can be expressed as implicit curves $T^m_\omega(\mathbf{p})=0$ with $m$ the curve label, each of which is the envelope of a single-parameter family of curves (See Appendix A for detail). Actually the CCE determined by Eq.~(\ref{CCE1}) are generally composed of $N$ disconnected simple closed curves labeled by $n=1,2,..N$ (e.g.~in $d$-wave cuprates the CCE consist of $N=4$ banana-shaped curves in certain energy range).
The $n$-th curve of the CCE can be represented by a parameter equation. For every parameter of the $n$-th curve, Eq.~(\ref{CCE2}) determines $N$ families of closed curves in the 2D $\mathbf{p}$ plane. Therefore there are totally $N\times N$ families of curves.
At last Eq.~(\ref{jacobi}) determines the envelope function of every family of curves, i.e.~$T^m_\omega(\mathbf{p})=0$.
Since the CCE are closed curves, the singularities $T^m_\omega(\mathbf{p})=0$ are also composed of closed curves.

These closed envelope curves then separate the $\mathbf{p}$ plane into different regions and serve as their boundaries. For any $\mathbf{p}$ in these regions the intersection points of Eq.~(\ref{CCE1}) and Eq.~(\ref{CCE2}) (i.e.~the real roots of these two coupled equations) are distinct, and the number of intersection points are even ($0,2,4,\dots$). Crossing the boundary (singularities) from one region to another, the number of distinct intersection points changes by $2$,
because on the boundaries there are degenerate roots (touching point) .

Eqs.~(\ref{JDOS}) and (\ref{jacobi}) also show the significant {\it choosing} effect of the singularities. Namely on the singular point $\mathbf{p}$ only the factors
$F_{\mathbf{k}_0,\mathbf{k}_0+\mathbf{p}}$ is highlighted. Therefore the variation of $F_{\mathbf{k},\mathbf{k}+\mathbf{p}}$ can be reflected in the JDOS data. This effect has been utilized to detect the gap symmetry of $d$-wave cuprates and iron-based superconductors via FT-STS~\cite{Hanaguri2009,Hanaguri2010}. As we show in the following section, FT-LDOS and JDOS share the same singularities. Therefore FT-LDOS can bear the same choosing effect as JDOS.


\section{Divergent behavior of JDOS and FT-LDOS}

It is easy to propose the idea of singularities from the aspect of JDOS as shown above. However the singular behavior at the vicinity of the singularities is difficult to obtain from Eqs.~(\ref{JDOS}), (\ref{CCE1}), (\ref{CCE2}) and (\ref{jacobi}). This problem and the relation between JDOS and FT-LDOS will be tackled in this section from another view. We suggest that the hot spots in the QPI pattern as observed by FT-STS are singularities of FT-LDOS, which can be captured by~\cite{QHan,DBZhang},
\begin{equation}
        S(\mathbf{p},\omega) = \int \frac{F_{\mathbf{k},\mathbf{k+p}}(\omega)}{\omega-E_\mathbf{k+p}} \delta(\omega-E_\mathbf{k}) dk_xdk_y,
\end{equation}
As in Eq.~(\ref{JDOSdef}) $F_{\mathbf{k},\mathbf{k+p}}(\omega)$ is responsible for the effects from nature of two or more components of quasiparticles and does not give rise to the singular behavior of FT-LDOS. In fact these singularities of FT-LDOS and JDOS are embedded in their elementary mathematical structures, i.e.~the rational and $\delta$ functions from real and imaginary parts of single-particle Green's function.
 Therefore we will temporarily omit $F_{\mathbf{k},\mathbf{k+p}}(\omega)$ and concentrate on the singularities intrinsically originating from the energy dispersion of the quasiparticles. $F_{\mathbf{k},\mathbf{k+p}}(\omega)$ will appear in the final results of FT-LDOS as a numerator, just like it does in Eq.~(\ref{JDOS}).

We first combine FT-LDOS with JDOS to form a complex response function as we did before~\cite{QHan,DBZhang},
\begin{equation}
    \begin{aligned}
        R(\mathbf{p},\omega) & = S(\mathbf{p},\omega) + i\pi J(\mathbf{p},\omega) \\
        &= \int \frac{1}{\omega-E_\mathbf{k+p}+i0^+} \delta(\omega-E_\mathbf{k}) dk_xdk_y,
    \end{aligned}
\end{equation}
with the help of which divergence behavior of and the relation between JDOS and FT-LDOS can be simultaneously obtained. The 2D integral is transformed to line integral along the CCE,
\begin{equation}
    \begin{aligned}
        R(\mathbf{p},\omega) =  \int_{E_\mathbf{k}=\omega}  \frac{1}{\omega-E_\mathbf{k+p}+i0^+} \frac{1}{|\nabla_\mathbf{k} E_\mathbf{k}|}dl,
    \end{aligned}
\end{equation}
For any given $\omega$ the CCE are generally composed of $N$ simple closed curves, each of which can be labeled by an integer $n=1,2,...,N$. As we assumed in the previous section the CCE do not cross any critical point. Therefore each closed curve can be parameterized by a periodic function. Here we use the arc-length parametrization followed by re-scaling of the period to $2\pi$. Therefore the explicit function of the curve can be written as $\mathbf{k}^n_\omega(t)=(f^n_\omega(t),g^n_\omega(t))$, where $n$ is the label of the closed curve, $f$, $g$ are regular and periodic functions with period $2\pi$. $dl_n(t)=\frac{P^n_\omega}{2\pi} dt'$ with $P^n_\omega$ the perimeter of the $n$-th closed curve.
After the parametrization $R(\mathbf{p},\omega)$ can be expressed as sum of line integral along every single closed curve,
\begin{equation}
    \label{line_int}
    \begin{aligned}
    R(\mathbf{p},\omega) = \sum_{n=1}^{N} \frac{P^n_\omega}{2\pi} \int_0^{2\pi} \frac{1}{\omega-E_{\mathbf{k}^n_\omega(t)+\mathbf{p}} +i0^+} \frac{dt}{v^n_\omega(t)},
    \end{aligned}
\end{equation}
where $v^n_\omega(t)=|\nabla_\mathbf{k} E_\mathbf{k}|_{\mathbf{k}=\mathbf{k}^n_\omega(t)}$. In the following derivation we will omit the index $n$ for brevity.

By observing Eq.~(\ref{line_int}) one can find that the integral might be singular if the denominator of the integrand becomes zero at certain $t_0$, that is
\begin{equation}
    \label{CCE2p}
    \omega-E_{\mathbf{k}_\omega(t_0)+\mathbf{p}}=0
\end{equation}
However if $t_0$ is a distinct root, i.e~its multiplicity is 1, it contributes a finite value to the integral. Actually Eq.~(\ref{CCE2p}) is the parameterized version of Eq.~(\ref{CCE2}) coupled with Eq.~(\ref{CCE1}). Therefore the distinct root is equivalent to distinct intersection points of
Eq.~(\ref{CCE1}) and (\ref{CCE2}). The integral is singular only if one of the real roots of Eq.~(\ref{CCE2p}) is degenerate, i.e.~
\begin{equation}
    \label{degenerate}
    \dot{E}_{\mathbf{k}_\omega(t_0)+\mathbf{p}}=\left.\frac{dE_{\mathbf{k}_\omega(t)+\mathbf{p}}}{dt}\right|_{t=t_0}=0.
\end{equation}
In fact, Eq.~(\ref{degenerate}) can be easily proved to be equivalent to Eq.~(\ref{jacobi}). Therefore $(\mathbf{p},\omega)$ is a singularity if Eq.~(\ref{CCE2p}) and (\ref{degenerate}) can be satisfied by certain real $t_0$.

For any singularity, $R(\mathbf{p},\omega)$ is infinite since the integrand has a second-order pole $t_0$  (see Appendix B for detail). What we are interested in is how it diverges to infinity when approaching to the singularities, namely the behavior of $R(\mathbf{p}+\bm{\delta},\omega)$ with $|\bm{\delta}|\sim0$. The result is, (see Appendix C for detail)
\begin{widetext}
\begin{equation}
    \label{sing}
    R(\mathbf{p}+\bm{\delta},\omega) \sim 
    - \frac{1}{|\mathbf{v}_{\mathbf{k}_0}||\mathbf{v}_{\mathbf{k}_0+\mathbf{p}}|}
    \frac{1}{|\bm{\delta}\cdot\bm{\varrho}_{\mathbf{k}_0+\mathbf{p},\mathbf{k}_0}|^{1/2}} 
    \left\{
    \begin{aligned}
        & \text{sgn}(\mathbf{n}_{\mathbf{k}_0+\mathbf{p}}\cdot\bm{\varrho}_{\mathbf{k}_0+\mathbf{p},\mathbf{k}_0}), & \bm{\delta}\cdot\bm{\varrho}_{\mathbf{k}_0+\mathbf{p},\mathbf{k}_0} > 0 & \\
        & i, & \bm{\delta}\cdot\bm{\varrho}_{\mathbf{k}_0+\mathbf{p},\mathbf{k}_0} < 0 &
    \end{aligned}
    \right.
    ,
\end{equation}
\end{widetext}
where we define the joint curvature,
\begin{equation}
    \label{j_curvature}
    \bm{\varrho}_{\mathbf{k}',\mathbf{k}}=\kappa_{\mathbf{k}'}\mathbf{n}_{\mathbf{k}'}- \kappa_{\mathbf{k}}\mathbf{n}_{\mathbf{k}}.
\end{equation}
Here $\kappa_\mathbf{k}$ is the curvature of the CCE at $\mathbf{k}$, whose definition is given in Appendix D.  $\mathbf{n}_\mathbf{k}=\mathbf{v}_\mathbf{k}/|\mathbf{v}_\mathbf{k}|$ is the direction vector of
the group velocity. $\mathbf{k}_0$ is the real solution of Eqs.~(\ref{CCE1}), (\ref{CCE2}) and (\ref{jacobi}). 

From Eq.~(\ref{sing}) we have the following results: (1) The so-called JDOS argument for $d$-wave superconductors is confirmed that the QPI pattern is proportional to $(|\mathbf{v}_i||\mathbf{v}_f|)^{-1}$ although it did not point out that the two velocities should be (anti)parallel as we show here. (2) There is a more important factor
governing the intensity of divergence, i.e.~the joint curvature $\bm{\varrho}_{\mathbf{k}_i,\mathbf{k}_f}$. (3) As real and imaginary parts of $R(\mathbf{p},\omega)$, 
FT-LDOS and JDOS share the same singularities and diverge complementarily in a inverse-square-root manner if any singularity is approached, unless the joint curvature is zero. (4) After a similar but tedious derivation as we did in Appendix B and C, we find that if the joint curvature of certain singularity is zero, the exponent of the power-law divergence is greater than $-2/3$. In fact the exponent is related to the multiplicity $M$ of the touching point by $M^{-1}-1$, with $M=2,3,...,\infty$. Larger value of $M$ means higher order of divergence and better nesting of the two CCE at $\mathbf{k}_0$ and $\mathbf{k}_0+\mathbf{p}$. And the infinite limit case corresponds to the perfect-nesting type of singularities.

\section{Discussion for some cases}

Near its singularities, the FT-LDOS $S(\mathbf{p},\omega)$ has stronger intensities and are reasonably easier to be observed by the FT-STS technique than regular ones.
Furthermore higher-order singularities, if they exist, are more robust than lower-order ones in the presence of various destruction effects on QPI, such as random impurity distribution, multiple impurity scattering, as well as short coherent length/time of quasiparticles. In the following we will show how to make use of Eq.~(\ref{sing}) to find the higher-order singularities.

\begin{figure}[ht]
\includegraphics[width=0.25\textwidth]{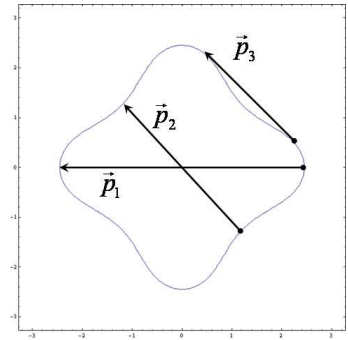}
\caption{Certain CCE for a toy model. By observation, we can see that the curvature of this CCE changes sign. $\mathbf{p}_{1,2,3}$ are singularities of
$S(\mathbf{p},\omega)$ because they connect points with their group velocities (anti)parallel. According to Eq.~(\ref{j_curvature}), only the joint curvature
of $\mathbf{p}_3$ is zero which leads to the fact that the order of $\mathbf{p}_3$ is higher than those of $\mathbf{p}_{1,2}$.}
\label{toy}
\end{figure}
If the CCE consist of one single curve whose curvature does not change sign, then the singularities of $S(\mathbf{p},\omega)$ are all $-1/2$ type although the magnitudes may be different depending on the detailed energy dispersion. However if the curvature of the CCE change sign, there must exist singularities higher-order than $-1/2$. Fig.~\ref{toy} show a CCE whose curvature changes sign. $\mathbf{p}_1$, $\mathbf{p}_2$ connect points with antiparallel velocities and same values of curvature. Therefore the joint curvature is nonzero according to Eq.~(\ref{j_curvature}). Therefore they are singularities of order $-1/2$. $\mathbf{p}_3$ is a singularity of
order $-2/3$ since its joint curvature is zero. 

\begin{figure}[ht]
\includegraphics[width=0.25\textwidth]{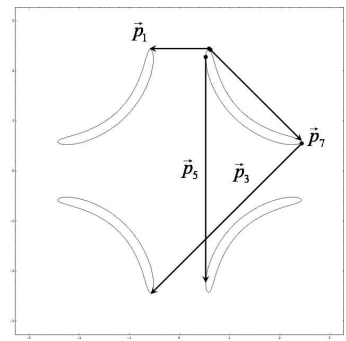}
\caption{Typical CCE of the $d$-wave superconductor with Bogoliubov quasiparticle energy lower than the spectral gap. Four banana-shaped close curves are shown
with higher-order singularities $\mathbf{p}_{1,3,5,7}$.}
\label{octet}
\end{figure}
In $d$-wave superconductors and iron-based superconductors with high anisotropy \cite{Hoffman,Allan}, some hot spots are resolved by FT-STS measurements among all the singularities, which are interpreted by the octet model. Some of them are obviously close to higher-order singularities. Fig.~\ref{octet} show the four vectors $\mathbf{p}_1$, $\mathbf{p}_3$, $\mathbf{p}_5$ and $\mathbf{p}_7$ of $d$-wave cuprates for illustration. $\mathbf{p}_{1,3,5,7}$ are all higher-order singularities because they connect points with parallel velocities and their joint curvatures are zero. Detailed examination shows that the existence of high-order $\mathbf{p}_{1,3,5}$ are guaranteed by the four-fold symmetry of the CCE of $d$-wave superconductors. On the other hand, the banana shape of the CCE, on which the curvature changes sign, is necessary for $\mathbf{p}_7$ to be higher ordered, just as $\mathbf{p}_3$ shown in Fig.~\ref{toy}. Compared with these four, $\mathbf{p}_{4}$ is obviously lower-order singular point whose joint curvature is nonzero, but the velocities are small so it still has higher intensity according to Eq.~(\ref{sing}) than other singularities except $\mathbf{p}_{1,3,5,7}$. Similar conclusion can also be applied to $\mathbf{p}_{2,6}$.

For comparison, the octet vector $\mathbf{q}_5$ are expected to connects the tips of the two banana-shaped CCE in the first and fourth quarter. Apparent difference between $\mathbf{p}_5$ and $\mathbf{q}_5$ can be clearly seen. $\mathbf{q}_5$ is actually a lower-order singularity because the velocities connected by it are antiparallel resulting in nonzero joint curvature. In fact the octet model is an approximate model whose validity is guaranteed by the strong isotropy of the energy dispersion, which leads to the fact that the octet vectors are close to the higher-order singularities.

\section{acknowledgement}

The authors thank Y.~Chen for helpful discussion. The work was supported by the Natural Science Foundation of China No.10674179, the RGC of Hong Kong under Grant No.HKU7055/09P, HKUST3/CRF/09, and the URC fund of HKU.

\section{Appendix A}

The CCE governed by Eq.~(\ref{CCE1}) are supposed to be composed of $N$ closed curves, each of which can be parameterized $k^i_x=f^i_\omega(c)$ and $k^i_y=g^i_\omega(c)$. $i=1,2,...,N$ the curve label and $c$ is the parameter. $(\dot{f}^i_\omega(c),\dot{g}^i_\omega(c))$ is the tangent vector of the $i$-th curve. Accordingly to Eq.~(\ref{CCE2}) we have the $N\times N$ one-parameter families of curves, each of which is given by
\begin{equation}
\begin{aligned}
    p^{i,j}_x(t,c)=f^i_{\omega}(t) - f^j_\omega(c), \\
    p^{i,j}_y(t,c)=g^i_{\omega}(t) - g^j_\omega(c),
\end{aligned}
   \ \ \  i,j=1,2,...,N
\end{equation}
whose envelope is determined by 
$$
\frac{\partial(p^{i,j}_x,p^{i,j}_y)}{\partial(t,c)}=\dot{f}^i_{\omega}(t)\dot{g}^j_{\omega}(c)-\dot{g}^i_{\omega}(t)\dot{f}^j_{\omega}(c)=0
$$ 
This indicates that the two tangent vectors are parallel, which is equivalent to Eq.~(\ref{jacobi}) because the corresponding group velocities are orthogonal to their corresponding tangent vectors.

\section{Appendix B}

To find the divergence behavior approaching to the singularities, we transform the above line integral to complex contour integral over counterclockwise unit circle by letting $z=e^{it}$, $\mathbf{K}_\omega(z)=(F_\omega(z),G_\omega(z))=(f_\omega(t),g_\omega(t))=\mathbf{k}_\omega(t)$, $V_\omega(z)=v_\omega(t)$,
\begin{equation}
    \begin{aligned}
        R(\mathbf{p},\omega)  = \frac{P_\omega}{2\pi i} \oint_{|z|=1} \frac{dz}{z V_\omega(z)[\omega-E_{\mathbf{K}_\omega(z)+\mathbf{p}}+i0^+]}.
    \end{aligned}
\end{equation}
Using the Cauchy residue theorem, we have
\begin{widetext}
\begin{equation}
    \label{residue}
    \begin{aligned}
    R(\mathbf{p},\omega) & = P_\omega \sum_k \text{Res} \{\frac{1}{zV_\omega(z)[\omega-E_{\mathbf{K}_\omega(z)+\mathbf{p}}+i0^+]},|z_k|<1 \}, \\
        & = C + P_\omega \sum_i [z_i V_\omega(z_i)]^{-1} \text{Res}\{ \frac{1}{\omega-E_{\mathbf{K}_\omega(z)+\mathbf{p}}+i0^+}, |z_i|<1 \},
    \end{aligned}
\end{equation}
\end{widetext}
where the sum in the first equality is over those $k$ for which $z_k$ is inside the unit circle. In the second equality we intentionally separate two parts of contributions. One part is denote by $C$ representing the sum of residues from poles of $[zV_\omega(z)]^{-1}$ which do not lead to singular contributions.
The other is a summation over the poles of $[\omega-\Xi_\omega(\mathbf{p},z)+i0^+]^{-1}$ which is infinitely close to the zeros of 
\begin{equation}
    \label{c_eqn}
    \omega-\Xi_\omega(\mathbf{p},z)=0.
\end{equation}
Eq.~(\ref{c_eqn}) is the complex version of Eq.~(\ref{CCE2p}) according to $z=e^{it}$. The real roots of Eq.~(\ref{CCE2p}) correspond to the complex zeros of Eq.~(\ref{c_eqn}) exactly on the unit circle. Complex roots of Eq.~(\ref{CCE2p}) with positive imaginary part give rise to complex zeros within the unit circle.
Since the complex zeros of Eq.~(\ref{c_eqn}) correspond to simple poles of Eq.~(\ref{residue}), they do not lead to singular contributions. The only source of divergence
originates from the degenerate real roots of Eq.~(\ref{CCE2p}). 

The existence of degenerate real roots demands that Eq.~(\ref{CCE2p}) and (\ref{degenerate}) can be simultaneously satisfied, which is realized if $(\mathbf{p},\omega)$ is a singular point. For this case we have from Eq.~(\ref{residue}),
\begin{widetext}
\begin{equation}
    R(\mathbf{p},\omega) \sim -\frac{P_\omega}{v_\omega(t_0)z_0} \sum_{z_\pm} \text{Res} \left[ \frac{1}{\frac{1}{2}\ddot{E}_{\mathbf{k}_\omega(t_0)+\mathbf{p}}(\frac{z-z_0}{iz_0})^2 -i0^+}, |z_\pm|<1 \right ].
\end{equation}
\end{widetext}
Clear $i0^+$ plays the role of determining only one of the two roots contributes to the integral. The result is
\begin{equation}
     R(\mathbf{p},\omega) \sim -\frac{P_\omega}{v_\omega(t_0)} \left[ \frac{\text{sgn}(\ddot{E}_{\mathbf{k}_\omega(t_0)+\mathbf{p}})+i}{|\ddot{E}_{\mathbf{k}_\omega(t_0)+\mathbf{p}}|^{1/2}} \right ] \frac{1}{\sqrt{0^+}},
\end{equation}
which indicate that on any singularity the value of $R$ is infinite. This result also gives clue on the fact that the singularities are
the common boundaries for the real and imaginary parts to diverge.

\section{Appendix C}

Here we give detailed derivation of the divergent behavior of $R(\mathbf{p},\omega)$ when any singulary is approached. If $(\mathbf{p},\omega)$ is a singularity, then the three coupled Eqs.~(\ref{CCE1}), (\ref{CCE2}), and (\ref{jacobi}) determine a touching point $\mathbf{k}_0$. Or equivalently Eq.~(\ref{CCE2p}) has a degenerate root $t_0$ satisfying (\ref{degenerate}). For any give $\omega$, when approaching to certain singularity $\mathbf{p}$ according to $R(\mathbf{p}+\bm{\delta},\omega)$ where $|\bm{\delta}|\sim 0$, we need to estimate how the degenerate root $t_0$ varies with $\bm{\delta}$. To leading order of $\bm{\delta}$ we have the pole equation,
\begin{equation}
\begin{aligned}
        0 &= \omega-E_{\mathbf{k}_\omega(t)+\mathbf{p}+\bm{\delta}} \\
         &\approx \omega-E_{\mathbf{k}_\omega(t)+\mathbf{p}}-\bm{\delta}\cdot\nabla E_{\mathbf{k}_\omega(t)+\mathbf{p}} \\
         &\approx  \frac{1}{2} \ddot{E}_{\mathbf{k}_\omega(t_0)+\mathbf{p}}(t-t_0)^2 + \bm{\delta}\cdot\nabla E_{\mathbf{k}_\omega(t_0)+\mathbf{p}}
\end{aligned}
\end{equation}
We have applied Eq.~(\ref{CCE2p}) and (\ref{degenerate}) in the derivation. From the above equation we find new roots at the vicinity of the degenerate root $t_0$,
\begin{equation}
    t_\pm=t_0 \pm (-\Delta)^{1/2}
\end{equation}
where
\begin{equation}
    \Delta\equiv\frac{2\bm{\delta}\cdot\nabla E_{\mathbf{k}_\omega(t_0)+\mathbf{p}}}{\ddot{E}_{\mathbf{k}_\omega(t_0)+\mathbf{p}}}
\end{equation}
If $\Delta>0$, $t_\pm=t_0\pm i|\Delta|^{1/2}$. 
The two CCE do not intersect at $\mathbf{k}_0=\mathbf{k}_\omega(t_0)$ any more. 
This indicates that we are approaching the singularity $\mathbf{p}$ from the region with less number of distinct intersection points. Transforming to the complex plane, $z_+=e^{it_+}$ is within the unit circle and therefore contribute to the integral according to Eq.~(\ref{residue}),
\begin{widetext}
\begin{equation}
\label{tmp1}
    \begin{aligned}
    R(\mathbf{p}+\bm{\delta},\omega) & \sim -\frac{P_\omega}{v_\omega(t_0)z_0} \text{Res} \left[ \frac{1}{\frac{1}{2}\ddot{E}_{\mathbf{k}_\omega(t_0)+\mathbf{p}}(\frac{z-z_0}{iz_0})^2
    +\nabla E_{\mathbf{k}_\omega(t_0)+\mathbf{p}}\cdot\bm{\delta}-i0^+}, z_+ \right ], \\
     & \sim -\frac{P_\omega}{v_\omega(t_0)} \left[ \frac{\text{sgn}(\ddot{E}_{\mathbf{k}_\omega(t_0)+\mathbf{p}})}{\sqrt{2\ddot{E}_{\mathbf{k}_\omega(t_0)+\mathbf{p}}\bm{\delta}\cdot\nabla E_{\mathbf{k}_\omega(t_0)+\mathbf{p}} }} \right ], \\
     & \sim -\frac{P_\omega}{|\nabla E_{\mathbf{k}_0}||\nabla E_{\mathbf{k}_0+\mathbf{p}}|} \left[ \frac{\text{sgn}(\ddot{E}_{\mathbf{k}_\omega(t_0)+\mathbf{p}})}{\sqrt{2\ddot{E}_{\mathbf{k}_\omega(t_0)+\mathbf{p}}\bm{\delta}\cdot \mathbf{n}_{\mathbf{k}_0+\mathbf{p}} }} \right ].
    \end{aligned}
\end{equation}
\end{widetext}
This results indicate that when approaching the singularity from the region of $\mathbf{p}$ where the real roots of Eqs.~(\ref{CCE1}) and (\ref{CCE2}) is less than
the other side of the singularities the real part of $R(\mathbf{p}+\bm{\delta},\omega)$, i.e~FT-LDOS, diverges according to $|\bm{\delta}|^{-1/2}$ where the imaginary
part, i.e.~JDOS does not.

If $\Delta<0$, $t_\pm=t_0\pm |\Delta|^{1/2}$ meaning that the degenerate real root $t_0$ now become two distinct real roots. This indicates that we are approaching the singularity $\mathbf{p}$ from the region with more number of distinct roots. For this case the infinitesimal imaginary number $i0^+$ plays the role of determining which one of the two poles $z_\pm=e^{it_\pm}$ lies in the unit circle in complex plane. The result is
\begin{equation}
\label{tmp2}
    \begin{aligned}
    R(\mathbf{p}+\bm{\delta},\omega)
     & \sim -\frac{P_\omega}{v_\omega(t_0)} \left[ \frac{i}{\sqrt{-2\ddot{E}_{\mathbf{k}_\omega(t_0)+\mathbf{p}}\bm{\delta}\cdot\nabla E_{\mathbf{k}_\omega(t_0)+\mathbf{p}} }} \right ],
    \end{aligned}
\end{equation}

Combining Eqs.~(\ref{tmp1}) and (\ref{tmp2}) together and making use of the result of $\ddot{E}$ given in Appendix D, we obtain Eq.~(\ref{sing}).

\section{Appendix D}

Considering that $t_0$ is a degenerate root satisfying Eqs.~(\ref{CCE2p}) and (\ref{degenerate}), we have
\begin{widetext}
\begin{equation}
    \begin{aligned}
    \ddot{E}_{\mathbf{k}_\omega(t_0)+\mathbf{p}} & = \left. \frac{d^2 E_{\mathbf{k}_\omega(t)+\mathbf{p}}}{dt^2} \right|_{t=t_0}  \\
    & = \left(\frac{P_\omega}{2\pi}\right)^2
    |\nabla E_{\mathbf{k}_{\omega(t_0)}+\mathbf{p}}| ( \kappa_{\mathbf{k}_\omega(t_0)+\mathbf{p}} - \mathbf{n}_{\mathbf{k}_\omega(t_0)+\mathbf{p}} \cdot \mathbf{n}_{\mathbf{k}_\omega(t_0)} \kappa_{\mathbf{k}_\omega(t_0)} ) \\
    & = \left(\frac{P_\omega}{2\pi}\right)^2
    |\nabla E_{\mathbf{k}_0+\mathbf{p}}| ( \kappa_{\mathbf{k}_0+\mathbf{p}} - \mathbf{n}_{\mathbf{k}_0+\mathbf{p}} \cdot \mathbf{n}_{\mathbf{k}_0} \kappa_{\mathbf{k}_0} ) \\
    & = \left(\frac{P_\omega}{2\pi}\right)^2 \nabla E_{\mathbf{k}_0+\mathbf{p}}\cdot\bm{\varrho}_{\mathbf{k}_0+\mathbf{p},\mathbf{k}_0} ,
    \end{aligned}
\end{equation}
\end{widetext}
where $\bm{\varrho}$ is the joint curvature defined according to Eq.~(\ref{j_curvature}) and $\mathbf{n}_\mathbf{k}=\frac{\nabla E_\mathbf{k}}{|\nabla E_\mathbf{k}|}$ the unit direction vector of the group velocity. Since $\mathbf{k}_0$ and $\mathbf{k}_0+\mathbf{p}$ satisfy Eqs.~(\ref{CCE1}), (\ref{CCE2}) and (\ref{jacobi}). Therefore $\mathbf{n}_{\mathbf{k}_0+\mathbf{p}} \cdot \mathbf{n}_{\mathbf{k}_0}=1$
if the group velocities are parallel and $-1$ if antiparallel. $\kappa_\mathbf{k}$ represents the curvature at point $\mathbf{k}$ on the CCE,
\begin{equation}
    \kappa_\mathbf{k} =  \frac{ (E_\mathbf{k}^y)^2 E_\mathbf{k}^{xx} -2E_\mathbf{k}^x E_\mathbf{k}^y E_\mathbf{k}^{xy}  + (E_\mathbf{k}^x)^2 E_\mathbf{k}^{yy}}
    {|\nabla E_\mathbf{k}|^{3/2}},
\end{equation}
where $E_\mathbf{k}^\alpha=\frac{\partial E_\mathbf{k}}{\partial k_\alpha}$, and 
$E_\mathbf{k}^{\alpha\beta}=\frac{\partial^2 E_\mathbf{k}}{\partial k_\alpha\partial k_\beta}$ with $\alpha, \beta=x,y$. Obviously if $\bm{\varrho}=0$, then $\ddot{E}=0$ and $t_0$ will be a degenerate root with multiplicity 3, which definitely gives rise to higher-order divergence.


\begin{thebibliography}{99}
\bibitem{note1}
In certain ideal case, e.g.~scattering off pure pair-breaking type of impurities, $F_{\mathbf{k},\mathbf{k}+\mathbf{p}}$ can become zero at certain momenta and destroy some singularities of JDOS related with those momenta~\cite{DBZhang}.
\end{thebibliography}
\end{document}